\documentclass[twocolumn,reprint,preprintnumbers,amsmath,amssymb,aps,prl,nofootinbib]{revtex4-1}

\usepackage{aas_macros}
\usepackage{graphicx}
\usepackage{dcolumn}
\usepackage{bm}
\usepackage{epsfig,psfrag,graphics,verbatim}
\usepackage{xcolor}
\usepackage{float}
\usepackage{amsmath,amssymb}
\usepackage{siunitx}
\usepackage{hyperref}
\usepackage{xfrac}

\hypersetup{
  colorlinks=true,
  linkcolor=blue,
  citecolor=blue,
  urlcolor=blue,
  hyperfootnotes=true
}

\newcommand{\beq}{\begin{eqnarray}}
\newcommand{\eeq}{\end{eqnarray}}

\newcommand{\gsim}{\lower.7ex\hbox{$\;\stackrel{\textstyle>}{\sim}\;$}}
\newcommand{\lsim}{\lower.7ex\hbox{$\;\stackrel{\textstyle<}{\sim}\;$}}

\newcommand{\be}{\begin{equation}} 
\newcommand{\ee}{\end{equation}}
\newcommand{\bea}{\begin{equation}\begin{aligned}}
\newcommand{\eea}{\end{aligned}\end{equation}}
\newcommand{\td}{{\rm d}}

\newcommand{\DM}{{\rm DM}}

\def\circa#1{\,\raise.3ex\hbox{$#1$\kern-.75em\lower1ex\hbox{$\sim$}}\,}

\begin{document}

\title{Dark Matter and the {\sc Xenon1T} electron recoil excess}

\author{Kristjan Kannike}
\author{Martti Raidal}
\affiliation{National Institute of Chemical Physics and Biophysics, R\"avala 10, Tallinn 10143, Estonia}

\author{Alessandro Strumia}
\affiliation{Dipartimento di Fisica dell'Universit{\`a} di Pisa, Italia}

\author{Daniele Teresi}
\affiliation{Dipartimento di Fisica dell'Universit{\`a} di Pisa and INFN, Italia}

\author{Hardi Veerm\"ae}
\affiliation{National Institute of Chemical Physics and Biophysics, R\"avala 10, Tallinn 10143, Estonia}

\date{\today}

\begin{abstract}
We show that the electron recoil excess around 2 keV claimed by the {\sc Xenon} collaboration can be fitted 
by DM or DM-like particles having a fast component with velocity of order $\sim 0.1$.
Those particles cannot be part of the cold DM halo of our Galaxy, so we speculate about their possible nature and origin,
such as fast moving DM sub-haloes, semi-annihilations of DM and relativistic axions produced by a nearby axion star.
Feasible new physics scenarios must  accommodate exotic DM dynamics and unusual DM properties. 
\end{abstract}

\maketitle

\section{Introduction}
The {\sc Xenon} collaboration reported results of searches for new physics 
with low-energy electronic recoil data recorded with the {\sc Xenon1T} detector~\cite{Aprile:2020tmw}. 
They claim an excess of events over the known backgrounds in the recoil energy $E_R$ range 1-7~keV, peaked around 2.4~keV. 
The local statistical significance is around  3-4$\sigma$, although part of the excess 
could be due to a small tritium background in the {\sc Xenon1T} detector~\cite{Aprile:2020tmw}.

The statistical significance of the excess is  $3.5\sigma$ when interpreted in terms of
axions~\cite{Peccei:1977hh,Weinberg:1977ma,Wilczek:1977pj} emitted by thermal processes in the Sun. 
This interpretation, considered by the {\sc Xenon} collaboration~\cite{Aprile:2020tmw},
depends on the axion-electron coupling $g_{ae}$, that dominates both the detection and the production in the Sun.  
Interestingly, the Sun emits such axions in the desired energy range with a spectrum that can fit the excess. 
Although the solar axion scenario has all the ingredients to explain the anomaly,  
the needed parameter space is excluded by stellar cooling bounds on the axion-electron coupling~\cite{Viaux:2013lha,Bertolami:2014wua,2016JCAP...07..036C,Battich:2016htm,Giannotti:2017hny}.  
Indeed, to fit the excess, the {\sc Xenon} collaboration requires $g_{ae} \approx 3.7\cdot 10^{-12}$ while 
stellar cooling constraints imply $g_{ae}\lsim 0.3\cdot 10^{-12}$. 
As the {\sc Xenon1T} rate scales as $g_{ae}^4$, bounds from cooling of hotter stars rule out this scenario quite convincingly.  
Neutrinos with a hypothetical magnetic moment are similarly excluded~\cite{Aprile:2020tmw, Bell:2006wi}.
Particles with small renormalizable interactions are less constrained by hotter stars~\cite{An:2013yua,An:2014twa}.
Still, this study demonstrates the need for a flux of fast particles in order to fit the data. 

Since DM exists, it is interesting to study whether the  {\sc Xenon1T} anomaly can be explained in some DM scenario. 
A narrow peak in the recoil energy $E_R$ can be provided by absorption of
a light bosonic DM particle~\cite{Aprile:2020tmw,1801725}
or by $\DM\,\DM\, e\to \DM \,e$ semi-absorption~\cite{Smirnov:2020zwf}.
Scatterings of DM particles provide a broader structure, but
cold DM particles are too slow, even when taking into account a Maxwellian-like tail of their velocity distribution around or above the cut-off due to the escape velocity from the Milky Way, $0.0015< v_{\rm esc}< 0.002$~\cite{Smith:2006ym} in natural units.
As we will see, the {\sc Xenon1T} excess needs a flux of fast (and possibly even relativistic)  particles. 



In this work we demonstrate that a flux of fast DM can provide a good fit to the {\sc Xenon1T} excess, and determine the necessary flux and velocity. 
Our results show that adding a free tritium abundance to the detector does not improve the fit. We later speculate about possible origins of such a fast DM component.

\section{Fast DM fit to {\sc Xenon1T} data}

We consider an elastic $\DM \,e\to \DM\, e'$ scattering between a DM particle with initial velocity $\vec v_\DM$
and an electron with initial velocity $\vec v_e$ that acquires final velocity $\vec v'_e$.
Assuming, for simplicity, that they are parallel and non-relativistic, the transferred recoil energy is
\bea \label{eq:estimate}
     E_R 
&   \equiv E_{e'} - E_e 
    = 2 \mu v_{\rm rel} v_{\rm CM}  \\
&   \simeq
\left\{\begin{array}{ll}
2 m_\DM v_e (v_\DM - v_e) & \hbox{for $m_\DM \ll m_e$},\\
2 m_e v_\DM(v_\DM - v_e)  & \hbox{for $m_\DM \gg m_e$},\\
\end{array}\right.
\eea
and the transferred momentum is
\bea
    q 
&   \equiv  m_\DM (v'_\DM - v_\DM ) 
    = -2\mu v_{\rm rel} \\
&   \simeq-
\left\{\begin{array}{ll}
2 m_\DM  (v_\DM -v_e) & \hbox{for $m_\DM \ll m_e$},\\
2 m_e  (v_\DM -v_e)  & \hbox{for $m_\DM \gg m_e$},\\
\end{array}\right.
\eea
where $v_{\rm CM} \equiv (m_e v_e + m_\DM v_\DM)/(m_e + m_\DM)$ is the center-of-mass velocity,
$v_{\rm rel} \equiv v_\DM - v_e$ is the relative velocity, 
and $\mu \equiv m_e m_\DM/(m_e + m_\DM)$ is the reduced mass. 
We see that the desired  $ E_R \sim 2.4\,{\rm keV}$ can be obtained 
for $m_\DM \gg m_e$ with $v_{\rm DM} \approx 0.1$ 
or for $m_\DM \ll m_e$ and faster DM, that becomes relativistic for $m_{\rm DM} \sim 0.1 m_e$.
Notice that $E_R  \simeq q v_{\rm CM}$ so that $q^2 \approx (40\,{\rm keV})^2$.

We validate the above estimates by performing a detailed computation taking into account the Xe atomic structure, along the lines of~\cite{Essig:2011nj,Roberts:2019chv}. In particular, we use the relativistic wave-functions for the ionization factor provided in~\cite{Roberts:2019chv}.

\begin{figure}[tb]
$$\includegraphics[width=0.45\textwidth]{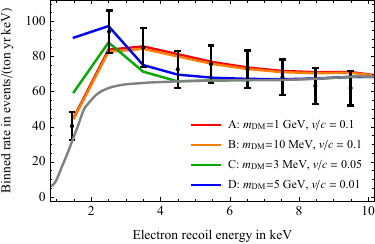} $$
\caption{\em Sample spectra for different values of DM mass and velocity. 
The gray curve is the background claimed by {\sc Xenon1T} (assuming negligible tritium contribution)
and the data points are shown in black. 
\label{fig:benchmarks}}
\end{figure}

For a fixed DM velocity $v_{\rm DM}$ (hereafter denoted by $v$ to simplify the notation), the differential cross-section is
\be 
	\frac{\td \sigma v}{\td E_R} = \frac{\sigma_e}{2 m_e v} \int_{q_-}^{q_+}\!\!\! a_0^2 \, q \, \td q |F(q)|^2 K(E_R,q),
\ee
where $\sigma_e$ is the free electron cross-section at fixed momentum transfer $q=1/a_0$, where $a_0=1/(\alpha m_e)$ is the Bohr radius. The limits of integration are
\be 
	q_\pm = m_{\rm DM} v \pm \sqrt{m_{\rm DM}^2 v^2 - 2 m_\DM E_R} .
\ee
We assume the DM form factor $F(q)=1$ obtained, {\it e.g.},~from heavy mediators. 
The atomic excitation factor $K(E_R,q)$ is taken from \cite{Roberts:2016xfw} and includes the relativistic corrections, relevant at large momentum exchange. For $E_R \sim $ keV recoil energies, the  excitation factor is dominated by the $n=3$ and $n=4$ atomic shells, the former starting at $E_R>1.17$ keV.
The differential rate is given by
\be 
	\frac{\td R}{\td E_R} = {n_T n_\DM}\frac{\td \sigma v}{\td E_R} \,,
\ee
where $n_T \simeq 4.2 \times 10^{27} / {\rm ton}$ is the number density of Xenon atoms, and $n_\DM$ is the number density of the fast DM component.
The rate depends on the product $n_\DM \sigma_e$, which we fit to the {\sc Xenon1T} excess. 
To compare the spectra with the {\sc Xenon1T} data, we smear them by a detector resolution $\sigma_{\rm det} = 0.45 \, \mathrm{keV}$ \cite{Aprile:2020yad}, approximated as constant, multiply by the efficiency given in~\cite{Aprile:2020tmw} and bin them as the available data in~\cite{Aprile:2020tmw}. We perform the fit both with negligible and free tritium abundance. In the latter case, the tritium signal shape is taken from~\cite{Aprile:2020yad} and its magnitude is fitted. 

Fig.~\ref{fig:benchmarks} compares the {\sc Xenon1T} data to sample spectra of the electron excess, computed for some values of the DM mass and velocity. 

\begin{figure}[t]
\begin{center}
\includegraphics[width=0.45\textwidth]{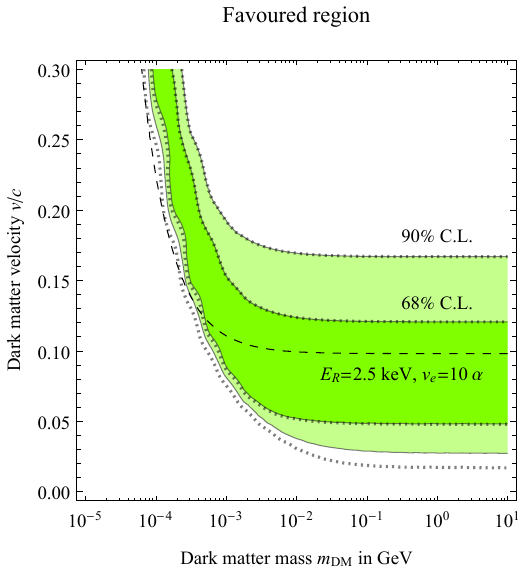}
\caption{\em Fit to the {\sc Xenon1T} excess as a function of the DM mass and velocity
assuming negligible tritium (continuous contours) and allowing for a free tritium abundance (dotted contours).
The numerical fit roughly follows the analytic estimate of eq.~\eqref{eq:estimate} (dashed curve).
\label{fig:chisq}}
\end{center}
\end{figure}

Fig.~\ref{fig:chisq} shows which values of these parameters best fit the energy spectrum of the excess.
We find that DM heavier than the electron with velocities $v_{\rm DM} \sim 0.1$ fits the excess well. 
On the other hand, lower masses do not provide sufficiently high electron recoil (unless the DM velocity is increased to relativistic values), whereas slower DM (even if heavier, to provide sufficient recoil) tends to give a too large signal in the first bin  1--2 keV. 
Allowing for a free amount of  tritium background (dotted contours in fig.~\ref{fig:chisq})
does not shift significantly the best-fit regions
because tritium reproduces the energy spectrum of the excess less well than fast DM.

\begin{figure}[tb]
\begin{center}
\includegraphics[width=0.45\textwidth]{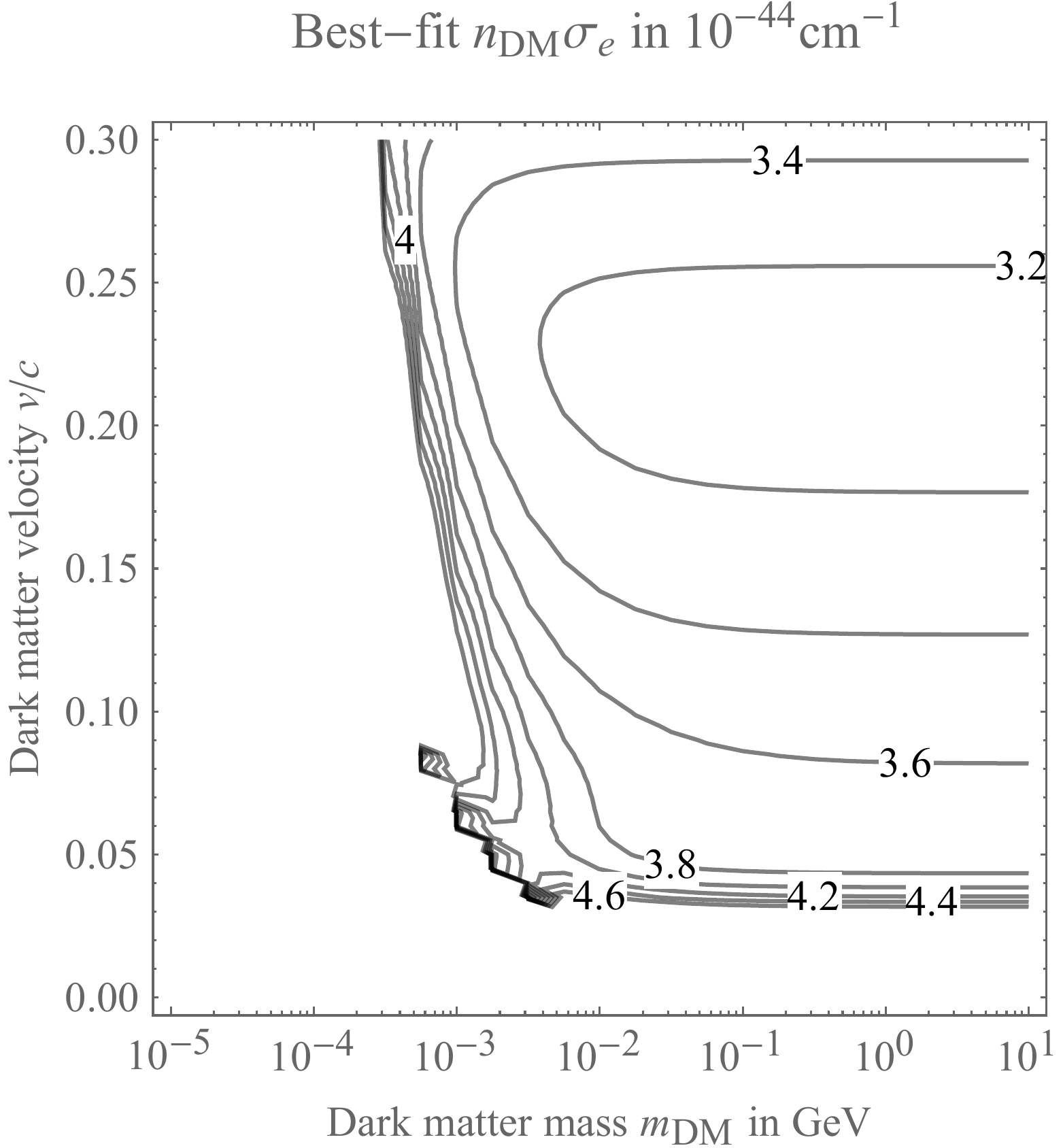}
\caption{\em Value of the DM number density (fast component) times cross section on electrons
that best fits the excess rate claimed by T as a function of the DM mass and velocity.
Regions that provide a good fit are shown in Fig.~\ref{fig:chisq}.
\label{fig:flux}}
\end{center}
\end{figure}

Fig.~\ref{fig:flux} shows the values of the number density of the fast DM component times its cross section on electrons
needed to reproduce the excess rate claimed by {\sc Xenon1T}.

We assumed that DM has a velocity distribution peaked at any given $v$ and constant $\sigma_{e}$.
If the cross section $\sigma_{e}$ were velocity-dependent, our fit applies with
$\sigma_{e}$ evaluated at $v$.
If the velocity distribution has a width, 
the fit still holds until it exceeds the {\sc Xenon} energy resolution.
Non-monochromatic velocity distributions produce broader signals.
In particular, the required population of fast DM particles cannot arise from a high-velocity tail of a broad
distribution (e.g.\ thermal), because such scenarios would be produce a too strong signal in the lower energy bin at 1--2 keV.

\section{Discussion and speculations}

We have seen that the {\sc Xenon1T} electron recoil excess can be interpreted as due to a flux of high velocity particles. 
Their velocities have to be so high that these particles cannot be gravitationally bound to the DM halo of our Galaxy.
Here we will speculate about possible physical origins of such flux of fast DM-like particles.
One needs to consider non-trivial DM dynamics, that must be consistent with all constraints.
For example, DM up-scattering by cosmic rays~\cite{Ema:2018bih,Yin:2018yjn} seems not to be consistent with other experiments.

One possibility is that the Earth is currently passing through a DM (sub-)halo that moves with a very high speed relative to us. 
The origin of such halo is, however, unclear, as the required velocities $v \circa{>} 0.05$ are an order of magnitude larger than the velocity dispersions in nearby rich galaxy clusters.



Another possibility is that a flux of fast DM is produced by semi-annihilation processes 
(see e.g.~\cite{Hambye:2008bq,DEramo:2010keq,Kamada:2017gfc,Kamada:2019wjo,Belanger:2012zr,Belanger:2014bga,Hektor:2019ote})
such as $\phi \phi\to\phi X$, where $\phi$ denotes the DM particle
and $X$ is extra particles.
In case $X$ has a negligible mass, a mono-energetic flux with $v_\DM =0.6$ is produced.
The speed can be different if, instead of $\phi$,  the final state contains a particle $\phi'$ with a different mass
(and possibly different interactions with electrons and SM particles).
A continuous spectrum is obtained if multiple particles $X$ are involved in the process.

DM heavier than $T/v_{\rm esc} \sim 1 \,{\rm GeV}$ can accumulate in the Sun or the Earth through elastic scattering with SM particles (see e.g. \cite{Garani:2017jcj} and \cite{Catena:2016tlv,Emken:2017erx}).
The resulting rate of semi-annihilation process in their centers is at most equal to the capture rate,
that is at most geometric.
In the most optimistic limit where all DM particles are captured
the equilibrium flux of fast DM particles from the Earth is $\Phi_\DM = \rho_\DM^{\rm slow} v_\DM^{\rm slow} /(8m_\DM)$, where $\rho_\DM^{\rm slow}\approx0.3\,{\rm GeV}/{\rm cm}^3 $ is the usual density of DM particles with $v_\DM^{\rm slow} \sim 10^{-3}$.
The {\sc Xenon} excess rate can be reproduced, for example, if the DM cross section on electrons
is a few orders of magnitude below  
$(2 R_E n_E)^{-1} \approx 3\times10^{-34}\,{\rm cm}^2$
the critical value for efficient capture by the Earth
(electron density $n_E \approx 4 N_A/{\rm cm}^3$, radius $R_E \approx 6400\,{\rm km}$)
consistently with experimental bounds~\cite{Essig:2012yx}.
Such cross section needs mediators with mass below the weak scale,
as electro-weak gauge invariance does not allow for dimension 5 operators.
On the other hand, the needed cross section is large enough that DM particles reflected from the Sun would thermalise and evaporate if lighter than about a GeV, providing 
a flux of fast DM~\cite{Chen:2020gcl}.

Similarly, one can consider a radioactive DM that slowly decays into energetic dark particles.
Another possibility is that DM contains structures similar to matter, with a lighter faster dark-electron
coupled by some dark photon to slower and heavier dark-nuclei, possibly in the form of dark-atoms (see e.g.~\cite{Clarke:2015gqw}).

\medskip

As a more exotic example, we consider a dark sector with a dissipative component that may contain dark stars~\cite{Kouvaris:2015rea,Chang:2018bgx}. Such stars can radiate relativistic DM particles and thus act as local sources. If, by coincidence, a nearby dark star exists, it could produce the required relativistic flux of fast DM. We remark that it may even be located in our solar system as the hypothetical Planet~9~\cite{2019PhR...805....1B,2016AJ....151...22B,Scholtz:2019csj}. 
For objects within the solar system, the signal is expected to be time dependent. 
Whether such exotic dark stars exist or not requires a dedicated study.

Finally, let us remark on a possible axion/dark photon solution to the {\sc Xenon1T}  result due to local sources.  While this is not directly related to the scattering of fast DM studied in this work,  it may replace the unviable solar axion solution considered in~\cite{Aprile:2020tmw}. In the axion DM scenario, DM may consist of axion stars~\cite{Kolb:1993zz, Levkov:2018kau, Eggemeier:2019jsu}.  As is well known, such scenarios require some hypothetical mechanism of axion star formation. The generic prediction of axion star dynamics is that they oscillate and emit both relativistic axions as well as photons in the form of radio bursts~\cite{Levkov:2018kau,Levkov:2020txo}.  The axion-electron coupling $g_{ae}$ (needed to explain the {\sc Xenon1T} results) can differ from the axion-photon coupling $g_{a\gamma}$ (which produces radio bursts and other axion signatures involving photons and electric and magnetic fields). So one can conceive  scenarios of axion stars that satisfy bounds from photon signals.  If a nearby axion star exists, it may be invisible and yet produce the required relativistic flux of axions without being excluded.

\section{Conclusions}
The excess in the electron recoil energy spectrum  around 2 keV
claimed by the {\sc Xenon1T} collaboration could be produced by fast DM or DM-like particles
hitting electrons with DM velocity $v \sim 0.1$.
A fast DM component is needed because the
cold DM with $v \sim10^{-3}$ recoiling on electrons produces an excess at lower energies.
This result persists the {\sc Xenon1T} is partly due to tritium background.
We speculated about possible exotic DM dynamics that produces the needed fast DM component.



{\small
\section*{Acknowledgements}
AS thanks Robert Foot, Christian Gross, Maxim Pospelov, Filippo Sala and Juri Smirnov for discussions.
DT thanks Dario Buttazzo and Paolo Panci for discussions and collaboration on related work. The work of AS and DT is supported by the ERC grant NEO-NAT. 
KK, MR and HV were supported by the Estonian Research Council grants PRG803 and PRG434, by the European Regional Development Fund and programme Mobilitas grants MOBTT86, MOBJD381, MOBTT5, MOBTP135, and by the EU through the European Regional Development Fund CoE program TK133 ``The Dark Side of the Universe".}

\bibliography{fast_DM}

\begin{thebibliography}{43}%
\makeatletter
\providecommand \@ifxundefined [1]{%
 \@ifx{#1\undefined}
}%
\providecommand \@ifnum [1]{%
 \ifnum #1\expandafter \@firstoftwo
 \else \expandafter \@secondoftwo
 \fi
}%
\providecommand \@ifx [1]{%
 \ifx #1\expandafter \@firstoftwo
 \else \expandafter \@secondoftwo
 \fi
}%
\providecommand \natexlab [1]{#1}%
\providecommand \enquote  [1]{``#1''}%
\providecommand \bibnamefont  [1]{#1}%
\providecommand \bibfnamefont [1]{#1}%
\providecommand \citenamefont [1]{#1}%
\providecommand \href@noop [0]{\@secondoftwo}%
\providecommand \href [0]{\begingroup \@sanitize@url \@href}%
\providecommand \@href[1]{\@@startlink{#1}\@@href}%
\providecommand \@@href[1]{\endgroup#1\@@endlink}%
\providecommand \@sanitize@url [0]{\catcode `\\12\catcode `\$12\catcode
  `\&12\catcode `\#12\catcode `\^12\catcode `\_12\catcode `\%12\relax}%
\providecommand \@@startlink[1]{}%
\providecommand \@@endlink[0]{}%
\providecommand \url  [0]{\begingroup\@sanitize@url \@url }%
\providecommand \@url [1]{\endgroup\@href {#1}{\urlprefix }}%
\providecommand \urlprefix  [0]{URL }%
\providecommand \Eprint [0]{\href }%
\providecommand \doibase [0]{http://dx.doi.org/}%
\providecommand \selectlanguage [0]{\@gobble}%
\providecommand \bibinfo  [0]{\@secondoftwo}%
\providecommand \bibfield  [0]{\@secondoftwo}%
\providecommand \translation [1]{[#1]}%
\providecommand \BibitemOpen [0]{}%
\providecommand \bibitemStop [0]{}%
\providecommand \bibitemNoStop [0]{.\EOS\space}%
\providecommand \EOS [0]{\spacefactor3000\relax}%
\providecommand \BibitemShut  [1]{\csname bibitem#1\endcsname}%
\let\auto@bib@innerbib\@empty
\bibitem [{\citenamefont {Aprile}\ \emph
  {et~al.}(2020{\natexlab{a}})\citenamefont {Aprile} \emph
  {et~al.}}]{Aprile:2020tmw}%
  \BibitemOpen
  \bibfield  {author} {\bibinfo {author} {\bibfnamefont {E.}~\bibnamefont
  {Aprile}} \emph {et~al.},\ }\href@noop {} {\  (\bibinfo {year}
  {2020}{\natexlab{a}})},\ \Eprint {http://arxiv.org/abs/2006.09721}
  {arXiv:2006.09721 [hep-ex]} \BibitemShut {NoStop}%
\bibitem [{\citenamefont {Peccei}\ and\ \citenamefont
  {Quinn}(1977)}]{Peccei:1977hh}%
  \BibitemOpen
  \bibfield  {author} {\bibinfo {author} {\bibfnamefont {R.}~\bibnamefont
  {Peccei}}\ and\ \bibinfo {author} {\bibfnamefont {H.~R.}\ \bibnamefont
  {Quinn}},\ }\href {\doibase 10.1103/PhysRevLett.38.1440} {\bibfield
  {journal} {\bibinfo  {journal} {Phys. Rev. Lett.}\ }\textbf {\bibinfo
  {volume} {38}},\ \bibinfo {pages} {1440} (\bibinfo {year}
  {1977})}\BibitemShut {NoStop}%
\bibitem [{\citenamefont {Weinberg}(1978)}]{Weinberg:1977ma}%
  \BibitemOpen
  \bibfield  {author} {\bibinfo {author} {\bibfnamefont {S.}~\bibnamefont
  {Weinberg}},\ }\href {\doibase 10.1103/PhysRevLett.40.223} {\bibfield
  {journal} {\bibinfo  {journal} {Phys. Rev. Lett.}\ }\textbf {\bibinfo
  {volume} {40}},\ \bibinfo {pages} {223} (\bibinfo {year} {1978})}\BibitemShut
  {NoStop}%
\bibitem [{\citenamefont {Wilczek}(1978)}]{Wilczek:1977pj}%
  \BibitemOpen
  \bibfield  {author} {\bibinfo {author} {\bibfnamefont {F.}~\bibnamefont
  {Wilczek}},\ }\href {\doibase 10.1103/PhysRevLett.40.279} {\bibfield
  {journal} {\bibinfo  {journal} {Phys. Rev. Lett.}\ }\textbf {\bibinfo
  {volume} {40}},\ \bibinfo {pages} {279} (\bibinfo {year} {1978})}\BibitemShut
  {NoStop}%
\bibitem [{\citenamefont {Viaux}\ \emph {et~al.}(2013)\citenamefont {Viaux},
  \citenamefont {Catelan}, \citenamefont {Stetson}, \citenamefont {Raffelt},
  \citenamefont {Redondo}, \citenamefont {Valcarce},\ and\ \citenamefont
  {Weiss}}]{Viaux:2013lha}%
  \BibitemOpen
  \bibfield  {author} {\bibinfo {author} {\bibfnamefont {N.}~\bibnamefont
  {Viaux}}, \bibinfo {author} {\bibfnamefont {M.}~\bibnamefont {Catelan}},
  \bibinfo {author} {\bibfnamefont {P.~B.}\ \bibnamefont {Stetson}}, \bibinfo
  {author} {\bibfnamefont {G.}~\bibnamefont {Raffelt}}, \bibinfo {author}
  {\bibfnamefont {J.}~\bibnamefont {Redondo}}, \bibinfo {author} {\bibfnamefont
  {A.~A.~R.}\ \bibnamefont {Valcarce}}, \ and\ \bibinfo {author} {\bibfnamefont
  {A.}~\bibnamefont {Weiss}},\ }\href {\doibase 10.1103/PhysRevLett.111.231301}
  {\bibfield  {journal} {\bibinfo  {journal} {Phys. Rev. Lett.}\ }\textbf
  {\bibinfo {volume} {111}},\ \bibinfo {pages} {231301} (\bibinfo {year}
  {2013})},\ \Eprint {http://arxiv.org/abs/1311.1669} {arXiv:1311.1669
  [astro-ph.SR]} \BibitemShut {NoStop}%
\bibitem [{\citenamefont {Miller~Bertolami}\ \emph {et~al.}(2014)\citenamefont
  {Miller~Bertolami}, \citenamefont {Melendez}, \citenamefont {Althaus},\ and\
  \citenamefont {Isern}}]{Bertolami:2014wua}%
  \BibitemOpen
  \bibfield  {author} {\bibinfo {author} {\bibfnamefont {M.~M.}\ \bibnamefont
  {Miller~Bertolami}}, \bibinfo {author} {\bibfnamefont {B.~E.}\ \bibnamefont
  {Melendez}}, \bibinfo {author} {\bibfnamefont {L.~G.}\ \bibnamefont
  {Althaus}}, \ and\ \bibinfo {author} {\bibfnamefont {J.}~\bibnamefont
  {Isern}},\ }\href {\doibase 10.1088/1475-7516/2014/10/069} {\bibfield
  {journal} {\bibinfo  {journal} {JCAP}\ }\textbf {\bibinfo {volume} {10}},\
  \bibinfo {pages} {069} (\bibinfo {year} {2014})},\ \Eprint
  {http://arxiv.org/abs/1406.7712} {arXiv:1406.7712 [hep-ph]} \BibitemShut
  {NoStop}%
\bibitem [{\citenamefont {{C{\'o}rsico}}\ \emph {et~al.}(2016)\citenamefont
  {{C{\'o}rsico}}, \citenamefont {{Romero}}, \citenamefont {{Althaus}},
  \citenamefont {{Garc{\'\i}a-Berro}}, \citenamefont {{Isern}}, \citenamefont
  {{Kepler}}, \citenamefont {{Miller Bertolami}}, \citenamefont {{Sullivan}},\
  and\ \citenamefont {{Chote}}}]{2016JCAP...07..036C}%
  \BibitemOpen
  \bibfield  {author} {\bibinfo {author} {\bibfnamefont {A.~H.}\ \bibnamefont
  {{C{\'o}rsico}}}, \bibinfo {author} {\bibfnamefont {A.~D.}\ \bibnamefont
  {{Romero}}}, \bibinfo {author} {\bibfnamefont {L.~r.~G.}\ \bibnamefont
  {{Althaus}}}, \bibinfo {author} {\bibfnamefont {E.}~\bibnamefont
  {{Garc{\'\i}a-Berro}}}, \bibinfo {author} {\bibfnamefont {J.}~\bibnamefont
  {{Isern}}}, \bibinfo {author} {\bibfnamefont {S.~O.}\ \bibnamefont
  {{Kepler}}}, \bibinfo {author} {\bibfnamefont {M.~M.}\ \bibnamefont {{Miller
  Bertolami}}}, \bibinfo {author} {\bibfnamefont {D.~J.}\ \bibnamefont
  {{Sullivan}}}, \ and\ \bibinfo {author} {\bibfnamefont {P.}~\bibnamefont
  {{Chote}}},\ }\href {\doibase 10.1088/1475-7516/2016/07/036} {\bibfield
  {journal} {\bibinfo  {journal} {\jcap}\ }\textbf {\bibinfo {volume} {2016}},\
  \bibinfo {eid} {036} (\bibinfo {year} {2016})},\ \Eprint
  {http://arxiv.org/abs/1605.06458} {arXiv:1605.06458 [astro-ph.SR]}
  \BibitemShut {NoStop}%
\bibitem [{\citenamefont {Battich}\ \emph {et~al.}(2016)\citenamefont
  {Battich}, \citenamefont {C{\'o}rsico}, \citenamefont {Althaus},
  \citenamefont {Miller~Bertolami},\ and\ \citenamefont
  {Bertolami}}]{Battich:2016htm}%
  \BibitemOpen
  \bibfield  {author} {\bibinfo {author} {\bibfnamefont {T.}~\bibnamefont
  {Battich}}, \bibinfo {author} {\bibfnamefont {A.~H.}\ \bibnamefont
  {C{\'o}rsico}}, \bibinfo {author} {\bibfnamefont {L.~G.}\ \bibnamefont
  {Althaus}}, \bibinfo {author} {\bibfnamefont {M.~M.}\ \bibnamefont
  {Miller~Bertolami}}, \ and\ \bibinfo {author} {\bibfnamefont
  {M.}~\bibnamefont {Bertolami}},\ }\href {\doibase
  10.1088/1475-7516/2016/08/062} {\bibfield  {journal} {\bibinfo  {journal}
  {JCAP}\ }\textbf {\bibinfo {volume} {08}},\ \bibinfo {pages} {062} (\bibinfo
  {year} {2016})},\ \Eprint {http://arxiv.org/abs/1605.07668} {arXiv:1605.07668
  [astro-ph.SR]} \BibitemShut {NoStop}%
\bibitem [{\citenamefont {Giannotti}\ \emph {et~al.}(2017)\citenamefont
  {Giannotti}, \citenamefont {Irastorza}, \citenamefont {Redondo},
  \citenamefont {Ringwald},\ and\ \citenamefont {Saikawa}}]{Giannotti:2017hny}%
  \BibitemOpen
  \bibfield  {author} {\bibinfo {author} {\bibfnamefont {M.}~\bibnamefont
  {Giannotti}}, \bibinfo {author} {\bibfnamefont {I.~G.}\ \bibnamefont
  {Irastorza}}, \bibinfo {author} {\bibfnamefont {J.}~\bibnamefont {Redondo}},
  \bibinfo {author} {\bibfnamefont {A.}~\bibnamefont {Ringwald}}, \ and\
  \bibinfo {author} {\bibfnamefont {K.}~\bibnamefont {Saikawa}},\ }\href
  {\doibase 10.1088/1475-7516/2017/10/010} {\bibfield  {journal} {\bibinfo
  {journal} {JCAP}\ }\textbf {\bibinfo {volume} {10}},\ \bibinfo {pages} {010}
  (\bibinfo {year} {2017})},\ \Eprint {http://arxiv.org/abs/1708.02111}
  {arXiv:1708.02111 [hep-ph]} \BibitemShut {NoStop}%
\bibitem [{\citenamefont {Bell}\ \emph {et~al.}(2006)\citenamefont {Bell},
  \citenamefont {Gorchtein}, \citenamefont {Ramsey-Musolf}, \citenamefont
  {Vogel},\ and\ \citenamefont {Wang}}]{Bell:2006wi}%
  \BibitemOpen
  \bibfield  {author} {\bibinfo {author} {\bibfnamefont {N.~F.}\ \bibnamefont
  {Bell}}, \bibinfo {author} {\bibfnamefont {M.}~\bibnamefont {Gorchtein}},
  \bibinfo {author} {\bibfnamefont {M.~J.}\ \bibnamefont {Ramsey-Musolf}},
  \bibinfo {author} {\bibfnamefont {P.}~\bibnamefont {Vogel}}, \ and\ \bibinfo
  {author} {\bibfnamefont {P.}~\bibnamefont {Wang}},\ }\href {\doibase
  10.1016/j.physletb.2006.09.055} {\bibfield  {journal} {\bibinfo  {journal}
  {Phys. Lett. B}\ }\textbf {\bibinfo {volume} {642}},\ \bibinfo {pages} {377}
  (\bibinfo {year} {2006})},\ \Eprint {http://arxiv.org/abs/hep-ph/0606248}
  {arXiv:hep-ph/0606248} \BibitemShut {NoStop}%
\bibitem [{\citenamefont {An}\ \emph {et~al.}(2013)\citenamefont {An},
  \citenamefont {Pospelov},\ and\ \citenamefont {Pradler}}]{An:2013yua}%
  \BibitemOpen
  \bibfield  {author} {\bibinfo {author} {\bibfnamefont {H.}~\bibnamefont
  {An}}, \bibinfo {author} {\bibfnamefont {M.}~\bibnamefont {Pospelov}}, \ and\
  \bibinfo {author} {\bibfnamefont {J.}~\bibnamefont {Pradler}},\ }\href
  {\doibase 10.1103/PhysRevLett.111.041302} {\bibfield  {journal} {\bibinfo
  {journal} {Phys. Rev. Lett.}\ }\textbf {\bibinfo {volume} {111}},\ \bibinfo
  {pages} {041302} (\bibinfo {year} {2013})},\ \Eprint
  {http://arxiv.org/abs/1304.3461} {arXiv:1304.3461 [hep-ph]} \BibitemShut
  {NoStop}%
\bibitem [{\citenamefont {An}\ \emph {et~al.}(2015)\citenamefont {An},
  \citenamefont {Pospelov}, \citenamefont {Pradler},\ and\ \citenamefont
  {Ritz}}]{An:2014twa}%
  \BibitemOpen
  \bibfield  {author} {\bibinfo {author} {\bibfnamefont {H.}~\bibnamefont
  {An}}, \bibinfo {author} {\bibfnamefont {M.}~\bibnamefont {Pospelov}},
  \bibinfo {author} {\bibfnamefont {J.}~\bibnamefont {Pradler}}, \ and\
  \bibinfo {author} {\bibfnamefont {A.}~\bibnamefont {Ritz}},\ }\href {\doibase
  10.1016/j.physletb.2015.06.018} {\bibfield  {journal} {\bibinfo  {journal}
  {Phys. Lett. B}\ }\textbf {\bibinfo {volume} {747}},\ \bibinfo {pages} {331}
  (\bibinfo {year} {2015})},\ \Eprint {http://arxiv.org/abs/1412.8378}
  {arXiv:1412.8378 [hep-ph]} \BibitemShut {NoStop}%
\bibitem [{\citenamefont {Takahashi}\ \emph {et~al.}(2020)\citenamefont
  {Takahashi}, \citenamefont {Yamada},\ and\ \citenamefont {Yin}}]{1801725}%
  \BibitemOpen
  \bibfield  {author} {\bibinfo {author} {\bibfnamefont {F.}~\bibnamefont
  {Takahashi}}, \bibinfo {author} {\bibfnamefont {M.}~\bibnamefont {Yamada}}, \
  and\ \bibinfo {author} {\bibfnamefont {W.}~\bibnamefont {Yin}},\ }\href@noop
  {} {\  (\bibinfo {year} {2020})},\ \Eprint {http://arxiv.org/abs/2006.10035}
  {arXiv:2006.10035 [hep-ph]} \BibitemShut {NoStop}%
\bibitem [{\citenamefont {Smirnov}\ and\ \citenamefont
  {Beacom}(2020)}]{Smirnov:2020zwf}%
  \BibitemOpen
  \bibfield  {author} {\bibinfo {author} {\bibfnamefont {J.}~\bibnamefont
  {Smirnov}}\ and\ \bibinfo {author} {\bibfnamefont {J.~F.}\ \bibnamefont
  {Beacom}},\ }\href@noop {} {\  (\bibinfo {year} {2020})},\ \Eprint
  {http://arxiv.org/abs/2002.04038} {arXiv:2002.04038 [hep-ph]} \BibitemShut
  {NoStop}%
\bibitem [{\citenamefont {Smith}\ \emph {et~al.}(2007)\citenamefont {Smith}
  \emph {et~al.}}]{Smith:2006ym}%
  \BibitemOpen
  \bibfield  {author} {\bibinfo {author} {\bibfnamefont {M.~C.}\ \bibnamefont
  {Smith}} \emph {et~al.},\ }\href {\doibase 10.1111/j.1365-2966.2007.11964.x}
  {\bibfield  {journal} {\bibinfo  {journal} {Mon. Not. Roy. Astron. Soc.}\
  }\textbf {\bibinfo {volume} {379}},\ \bibinfo {pages} {755} (\bibinfo {year}
  {2007})},\ \Eprint {http://arxiv.org/abs/astro-ph/0611671}
  {arXiv:astro-ph/0611671} \BibitemShut {NoStop}%
\bibitem [{\citenamefont {Essig}\ \emph
  {et~al.}(2012{\natexlab{a}})\citenamefont {Essig}, \citenamefont {Mardon},\
  and\ \citenamefont {Volansky}}]{Essig:2011nj}%
  \BibitemOpen
  \bibfield  {author} {\bibinfo {author} {\bibfnamefont {R.}~\bibnamefont
  {Essig}}, \bibinfo {author} {\bibfnamefont {J.}~\bibnamefont {Mardon}}, \
  and\ \bibinfo {author} {\bibfnamefont {T.}~\bibnamefont {Volansky}},\ }\href
  {\doibase 10.1103/PhysRevD.85.076007} {\bibfield  {journal} {\bibinfo
  {journal} {Phys. Rev. D}\ }\textbf {\bibinfo {volume} {85}},\ \bibinfo
  {pages} {076007} (\bibinfo {year} {2012}{\natexlab{a}})},\ \Eprint
  {http://arxiv.org/abs/1108.5383} {arXiv:1108.5383 [hep-ph]} \BibitemShut
  {NoStop}%
\bibitem [{\citenamefont {Roberts}\ and\ \citenamefont
  {Flambaum}(2019)}]{Roberts:2019chv}%
  \BibitemOpen
  \bibfield  {author} {\bibinfo {author} {\bibfnamefont {B.}~\bibnamefont
  {Roberts}}\ and\ \bibinfo {author} {\bibfnamefont {V.}~\bibnamefont
  {Flambaum}},\ }\href {\doibase 10.1103/PhysRevD.100.063017} {\bibfield
  {journal} {\bibinfo  {journal} {Phys. Rev. D}\ }\textbf {\bibinfo {volume}
  {100}},\ \bibinfo {pages} {063017} (\bibinfo {year} {2019})},\ \Eprint
  {http://arxiv.org/abs/1904.07127} {arXiv:1904.07127 [hep-ph]} \BibitemShut
  {NoStop}%
\bibitem [{\citenamefont {Roberts}\ \emph {et~al.}(2016)\citenamefont
  {Roberts}, \citenamefont {Dzuba}, \citenamefont {Flambaum}, \citenamefont
  {Pospelov},\ and\ \citenamefont {Stadnik}}]{Roberts:2016xfw}%
  \BibitemOpen
  \bibfield  {author} {\bibinfo {author} {\bibfnamefont {B.}~\bibnamefont
  {Roberts}}, \bibinfo {author} {\bibfnamefont {V.}~\bibnamefont {Dzuba}},
  \bibinfo {author} {\bibfnamefont {V.}~\bibnamefont {Flambaum}}, \bibinfo
  {author} {\bibfnamefont {M.}~\bibnamefont {Pospelov}}, \ and\ \bibinfo
  {author} {\bibfnamefont {Y.}~\bibnamefont {Stadnik}},\ }\href {\doibase
  10.1103/PhysRevD.93.115037} {\bibfield  {journal} {\bibinfo  {journal} {Phys.
  Rev. D}\ }\textbf {\bibinfo {volume} {93}},\ \bibinfo {pages} {115037}
  (\bibinfo {year} {2016})},\ \Eprint {http://arxiv.org/abs/1604.04559}
  {arXiv:1604.04559 [hep-ph]} \BibitemShut {NoStop}%
\bibitem [{\citenamefont {Aprile}\ \emph
  {et~al.}(2020{\natexlab{b}})\citenamefont {Aprile} \emph
  {et~al.}}]{Aprile:2020yad}%
  \BibitemOpen
  \bibfield  {author} {\bibinfo {author} {\bibfnamefont {E.}~\bibnamefont
  {Aprile}} \emph {et~al.} (\bibinfo {collaboration} {XENON}),\ }\href@noop {}
  {\  (\bibinfo {year} {2020}{\natexlab{b}})},\ \Eprint
  {http://arxiv.org/abs/2003.03825} {arXiv:2003.03825 [physics.ins-det]}
  \BibitemShut {NoStop}%
\bibitem [{\citenamefont {Ema}\ \emph {et~al.}(2019)\citenamefont {Ema},
  \citenamefont {Sala},\ and\ \citenamefont {Sato}}]{Ema:2018bih}%
  \BibitemOpen
  \bibfield  {author} {\bibinfo {author} {\bibfnamefont {Y.}~\bibnamefont
  {Ema}}, \bibinfo {author} {\bibfnamefont {F.}~\bibnamefont {Sala}}, \ and\
  \bibinfo {author} {\bibfnamefont {R.}~\bibnamefont {Sato}},\ }\href {\doibase
  10.1103/PhysRevLett.122.181802} {\bibfield  {journal} {\bibinfo  {journal}
  {Phys. Rev. Lett.}\ }\textbf {\bibinfo {volume} {122}},\ \bibinfo {pages}
  {181802} (\bibinfo {year} {2019})},\ \Eprint
  {http://arxiv.org/abs/1811.00520} {arXiv:1811.00520 [hep-ph]} \BibitemShut
  {NoStop}%
\bibitem [{\citenamefont {Yin}(2019)}]{Yin:2018yjn}%
  \BibitemOpen
  \bibfield  {author} {\bibinfo {author} {\bibfnamefont {W.}~\bibnamefont
  {Yin}},\ }\href {\doibase 10.1051/epjconf/201920804003} {\bibfield  {journal}
  {\bibinfo  {journal} {EPJ Web Conf.}\ }\textbf {\bibinfo {volume} {208}},\
  \bibinfo {pages} {04003} (\bibinfo {year} {2019})},\ \Eprint
  {http://arxiv.org/abs/1809.08610} {arXiv:1809.08610 [hep-ph]} \BibitemShut
  {NoStop}%
\bibitem [{\citenamefont {Hambye}(2009)}]{Hambye:2008bq}%
  \BibitemOpen
  \bibfield  {author} {\bibinfo {author} {\bibfnamefont {T.}~\bibnamefont
  {Hambye}},\ }\href {\doibase 10.1088/1126-6708/2009/01/028} {\bibfield
  {journal} {\bibinfo  {journal} {JHEP}\ }\textbf {\bibinfo {volume} {01}},\
  \bibinfo {pages} {028} (\bibinfo {year} {2009})},\ \Eprint
  {http://arxiv.org/abs/0811.0172} {arXiv:0811.0172 [hep-ph]} \BibitemShut
  {NoStop}%
\bibitem [{\citenamefont {D'Eramo}\ and\ \citenamefont
  {Thaler}(2010)}]{DEramo:2010keq}%
  \BibitemOpen
  \bibfield  {author} {\bibinfo {author} {\bibfnamefont {F.}~\bibnamefont
  {D'Eramo}}\ and\ \bibinfo {author} {\bibfnamefont {J.}~\bibnamefont
  {Thaler}},\ }\href {\doibase 10.1007/JHEP06(2010)109} {\bibfield  {journal}
  {\bibinfo  {journal} {JHEP}\ }\textbf {\bibinfo {volume} {06}},\ \bibinfo
  {pages} {109} (\bibinfo {year} {2010})},\ \Eprint
  {http://arxiv.org/abs/1003.5912} {arXiv:1003.5912 [hep-ph]} \BibitemShut
  {NoStop}%
\bibitem [{\citenamefont {Kamada}\ \emph {et~al.}(2018)\citenamefont {Kamada},
  \citenamefont {Kim}, \citenamefont {Kim},\ and\ \citenamefont
  {Sekiguchi}}]{Kamada:2017gfc}%
  \BibitemOpen
  \bibfield  {author} {\bibinfo {author} {\bibfnamefont {A.}~\bibnamefont
  {Kamada}}, \bibinfo {author} {\bibfnamefont {H.~J.}\ \bibnamefont {Kim}},
  \bibinfo {author} {\bibfnamefont {H.}~\bibnamefont {Kim}}, \ and\ \bibinfo
  {author} {\bibfnamefont {T.}~\bibnamefont {Sekiguchi}},\ }\href {\doibase
  10.1103/PhysRevLett.120.131802} {\bibfield  {journal} {\bibinfo  {journal}
  {Phys. Rev. Lett.}\ }\textbf {\bibinfo {volume} {120}},\ \bibinfo {pages}
  {131802} (\bibinfo {year} {2018})},\ \Eprint
  {http://arxiv.org/abs/1707.09238} {arXiv:1707.09238 [hep-ph]} \BibitemShut
  {NoStop}%
\bibitem [{\citenamefont {Kamada}\ and\ \citenamefont
  {Kim}(2019)}]{Kamada:2019wjo}%
  \BibitemOpen
  \bibfield  {author} {\bibinfo {author} {\bibfnamefont {A.}~\bibnamefont
  {Kamada}}\ and\ \bibinfo {author} {\bibfnamefont {H.~J.}\ \bibnamefont
  {Kim}},\ }\href@noop {} {\  (\bibinfo {year} {2019})},\ \Eprint
  {http://arxiv.org/abs/1911.09717} {arXiv:1911.09717 [hep-ph]} \BibitemShut
  {NoStop}%
\bibitem [{\citenamefont {Belanger}\ \emph {et~al.}(2013)\citenamefont
  {Belanger}, \citenamefont {Kannike}, \citenamefont {Pukhov},\ and\
  \citenamefont {Raidal}}]{Belanger:2012zr}%
  \BibitemOpen
  \bibfield  {author} {\bibinfo {author} {\bibfnamefont {G.}~\bibnamefont
  {Belanger}}, \bibinfo {author} {\bibfnamefont {K.}~\bibnamefont {Kannike}},
  \bibinfo {author} {\bibfnamefont {A.}~\bibnamefont {Pukhov}}, \ and\ \bibinfo
  {author} {\bibfnamefont {M.}~\bibnamefont {Raidal}},\ }\href {\doibase
  10.1088/1475-7516/2013/01/022} {\bibfield  {journal} {\bibinfo  {journal}
  {JCAP}\ }\textbf {\bibinfo {volume} {01}},\ \bibinfo {pages} {022} (\bibinfo
  {year} {2013})},\ \Eprint {http://arxiv.org/abs/1211.1014} {arXiv:1211.1014
  [hep-ph]} \BibitemShut {NoStop}%
\bibitem [{\citenamefont {B{\'e}langer}\ \emph {et~al.}(2014)\citenamefont
  {B{\'e}langer}, \citenamefont {Kannike}, \citenamefont {Pukhov},\ and\
  \citenamefont {Raidal}}]{Belanger:2014bga}%
  \BibitemOpen
  \bibfield  {author} {\bibinfo {author} {\bibfnamefont {G.}~\bibnamefont
  {B{\'e}langer}}, \bibinfo {author} {\bibfnamefont {K.}~\bibnamefont
  {Kannike}}, \bibinfo {author} {\bibfnamefont {A.}~\bibnamefont {Pukhov}}, \
  and\ \bibinfo {author} {\bibfnamefont {M.}~\bibnamefont {Raidal}},\ }\href
  {\doibase 10.1088/1475-7516/2014/06/021} {\bibfield  {journal} {\bibinfo
  {journal} {JCAP}\ }\textbf {\bibinfo {volume} {06}},\ \bibinfo {pages} {021}
  (\bibinfo {year} {2014})},\ \Eprint {http://arxiv.org/abs/1403.4960}
  {arXiv:1403.4960 [hep-ph]} \BibitemShut {NoStop}%
\bibitem [{\citenamefont {Hektor}\ \emph {et~al.}(2019)\citenamefont {Hektor},
  \citenamefont {Hryczuk},\ and\ \citenamefont {Kannike}}]{Hektor:2019ote}%
  \BibitemOpen
  \bibfield  {author} {\bibinfo {author} {\bibfnamefont {A.}~\bibnamefont
  {Hektor}}, \bibinfo {author} {\bibfnamefont {A.}~\bibnamefont {Hryczuk}}, \
  and\ \bibinfo {author} {\bibfnamefont {K.}~\bibnamefont {Kannike}},\ }\href
  {\doibase 10.1007/JHEP03(2019)204} {\bibfield  {journal} {\bibinfo  {journal}
  {JHEP}\ }\textbf {\bibinfo {volume} {03}},\ \bibinfo {pages} {204} (\bibinfo
  {year} {2019})},\ \Eprint {http://arxiv.org/abs/1901.08074} {arXiv:1901.08074
  [hep-ph]} \BibitemShut {NoStop}%
\bibitem [{\citenamefont {Garani}\ and\ \citenamefont
  {Palomares-Ruiz}(2017)}]{Garani:2017jcj}%
  \BibitemOpen
  \bibfield  {author} {\bibinfo {author} {\bibfnamefont {R.}~\bibnamefont
  {Garani}}\ and\ \bibinfo {author} {\bibfnamefont {S.}~\bibnamefont
  {Palomares-Ruiz}},\ }\href {\doibase 10.1088/1475-7516/2017/05/007}
  {\bibfield  {journal} {\bibinfo  {journal} {JCAP}\ }\textbf {\bibinfo
  {volume} {05}},\ \bibinfo {pages} {007} (\bibinfo {year} {2017})},\ \Eprint
  {http://arxiv.org/abs/1702.02768} {arXiv:1702.02768 [hep-ph]} \BibitemShut
  {NoStop}%
\bibitem [{\citenamefont {Catena}\ and\ \citenamefont
  {Kouvaris}(2017)}]{Catena:2016tlv}%
  \BibitemOpen
  \bibfield  {author} {\bibinfo {author} {\bibfnamefont {R.}~\bibnamefont
  {Catena}}\ and\ \bibinfo {author} {\bibfnamefont {C.}~\bibnamefont
  {Kouvaris}},\ }\href {\doibase 10.1103/PhysRevD.96.063012} {\bibfield
  {journal} {\bibinfo  {journal} {Phys. Rev. D}\ }\textbf {\bibinfo {volume}
  {96}},\ \bibinfo {pages} {063012} (\bibinfo {year} {2017})},\ \Eprint
  {http://arxiv.org/abs/1608.07296} {arXiv:1608.07296 [astro-ph.CO]}
  \BibitemShut {NoStop}%
\bibitem [{\citenamefont {Emken}\ \emph {et~al.}(2017)\citenamefont {Emken},
  \citenamefont {Kouvaris},\ and\ \citenamefont {Shoemaker}}]{Emken:2017erx}%
  \BibitemOpen
  \bibfield  {author} {\bibinfo {author} {\bibfnamefont {T.}~\bibnamefont
  {Emken}}, \bibinfo {author} {\bibfnamefont {C.}~\bibnamefont {Kouvaris}}, \
  and\ \bibinfo {author} {\bibfnamefont {I.~M.}\ \bibnamefont {Shoemaker}},\
  }\href {\doibase 10.1103/PhysRevD.96.015018} {\bibfield  {journal} {\bibinfo
  {journal} {Phys. Rev. D}\ }\textbf {\bibinfo {volume} {96}},\ \bibinfo
  {pages} {015018} (\bibinfo {year} {2017})},\ \Eprint
  {http://arxiv.org/abs/1702.07750} {arXiv:1702.07750 [hep-ph]} \BibitemShut
  {NoStop}%
\bibitem [{\citenamefont {Essig}\ \emph
  {et~al.}(2012{\natexlab{b}})\citenamefont {Essig}, \citenamefont
  {Manalaysay}, \citenamefont {Mardon}, \citenamefont {Sorensen},\ and\
  \citenamefont {Volansky}}]{Essig:2012yx}%
  \BibitemOpen
  \bibfield  {author} {\bibinfo {author} {\bibfnamefont {R.}~\bibnamefont
  {Essig}}, \bibinfo {author} {\bibfnamefont {A.}~\bibnamefont {Manalaysay}},
  \bibinfo {author} {\bibfnamefont {J.}~\bibnamefont {Mardon}}, \bibinfo
  {author} {\bibfnamefont {P.}~\bibnamefont {Sorensen}}, \ and\ \bibinfo
  {author} {\bibfnamefont {T.}~\bibnamefont {Volansky}},\ }\href {\doibase
  10.1103/PhysRevLett.109.021301} {\bibfield  {journal} {\bibinfo  {journal}
  {Phys. Rev. Lett.}\ }\textbf {\bibinfo {volume} {109}},\ \bibinfo {pages}
  {021301} (\bibinfo {year} {2012}{\natexlab{b}})},\ \Eprint
  {http://arxiv.org/abs/1206.2644} {arXiv:1206.2644 [astro-ph.CO]} \BibitemShut
  {NoStop}%
\bibitem [{\citenamefont {Chen}\ \emph {et~al.}(2020)\citenamefont {Chen},
  \citenamefont {Shu}, \citenamefont {Xue}, \citenamefont {Yuan},\ and\
  \citenamefont {Yuan}}]{Chen:2020gcl}%
  \BibitemOpen
  \bibfield  {author} {\bibinfo {author} {\bibfnamefont {Y.}~\bibnamefont
  {Chen}}, \bibinfo {author} {\bibfnamefont {J.}~\bibnamefont {Shu}}, \bibinfo
  {author} {\bibfnamefont {X.}~\bibnamefont {Xue}}, \bibinfo {author}
  {\bibfnamefont {G.}~\bibnamefont {Yuan}}, \ and\ \bibinfo {author}
  {\bibfnamefont {Q.}~\bibnamefont {Yuan}},\ }\href@noop {} {\  (\bibinfo
  {year} {2020})},\ \Eprint {http://arxiv.org/abs/2006.12447} {arXiv:2006.12447
  [hep-ph]} \BibitemShut {NoStop}%
\bibitem [{\citenamefont {Clarke}\ and\ \citenamefont
  {Foot}(2016)}]{Clarke:2015gqw}%
  \BibitemOpen
  \bibfield  {author} {\bibinfo {author} {\bibfnamefont {J.~D.}\ \bibnamefont
  {Clarke}}\ and\ \bibinfo {author} {\bibfnamefont {R.}~\bibnamefont {Foot}},\
  }\href {\doibase 10.1088/1475-7516/2016/01/029} {\bibfield  {journal}
  {\bibinfo  {journal} {JCAP}\ }\textbf {\bibinfo {volume} {01}},\ \bibinfo
  {pages} {029} (\bibinfo {year} {2016})},\ \Eprint
  {http://arxiv.org/abs/1512.06471} {arXiv:1512.06471 [astro-ph.GA]}
  \BibitemShut {NoStop}%
\bibitem [{\citenamefont {Kouvaris}\ and\ \citenamefont
  {Nielsen}(2015)}]{Kouvaris:2015rea}%
  \BibitemOpen
  \bibfield  {author} {\bibinfo {author} {\bibfnamefont {C.}~\bibnamefont
  {Kouvaris}}\ and\ \bibinfo {author} {\bibfnamefont {N.~G.}\ \bibnamefont
  {Nielsen}},\ }\href {\doibase 10.1103/PhysRevD.92.063526} {\bibfield
  {journal} {\bibinfo  {journal} {Phys. Rev. D}\ }\textbf {\bibinfo {volume}
  {92}},\ \bibinfo {pages} {063526} (\bibinfo {year} {2015})},\ \Eprint
  {http://arxiv.org/abs/1507.00959} {arXiv:1507.00959 [hep-ph]} \BibitemShut
  {NoStop}%
\bibitem [{\citenamefont {Chang}\ \emph {et~al.}(2019)\citenamefont {Chang},
  \citenamefont {Egana-Ugrinovic}, \citenamefont {Essig},\ and\ \citenamefont
  {Kouvaris}}]{Chang:2018bgx}%
  \BibitemOpen
  \bibfield  {author} {\bibinfo {author} {\bibfnamefont {J.~H.}\ \bibnamefont
  {Chang}}, \bibinfo {author} {\bibfnamefont {D.}~\bibnamefont
  {Egana-Ugrinovic}}, \bibinfo {author} {\bibfnamefont {R.}~\bibnamefont
  {Essig}}, \ and\ \bibinfo {author} {\bibfnamefont {C.}~\bibnamefont
  {Kouvaris}},\ }\href {\doibase 10.1088/1475-7516/2019/03/036} {\bibfield
  {journal} {\bibinfo  {journal} {JCAP}\ }\textbf {\bibinfo {volume} {03}},\
  \bibinfo {pages} {036} (\bibinfo {year} {2019})},\ \Eprint
  {http://arxiv.org/abs/1812.07000} {arXiv:1812.07000 [hep-ph]} \BibitemShut
  {NoStop}%
\bibitem [{\citenamefont {{Batygin}}\ \emph {et~al.}(2019)\citenamefont
  {{Batygin}}, \citenamefont {{Adams}}, \citenamefont {{Brown}},\ and\
  \citenamefont {{Becker}}}]{2019PhR...805....1B}%
  \BibitemOpen
  \bibfield  {author} {\bibinfo {author} {\bibfnamefont {K.}~\bibnamefont
  {{Batygin}}}, \bibinfo {author} {\bibfnamefont {F.~C.}\ \bibnamefont
  {{Adams}}}, \bibinfo {author} {\bibfnamefont {M.~E.}\ \bibnamefont
  {{Brown}}}, \ and\ \bibinfo {author} {\bibfnamefont {J.~C.}\ \bibnamefont
  {{Becker}}},\ }\href {\doibase 10.1016/j.physrep.2019.01.009} {\bibfield
  {journal} {\bibinfo  {journal} {\physrep}\ }\textbf {\bibinfo {volume}
  {805}},\ \bibinfo {pages} {1} (\bibinfo {year} {2019})},\ \Eprint
  {http://arxiv.org/abs/1902.10103} {arXiv:1902.10103 [astro-ph.EP]}
  \BibitemShut {NoStop}%
\bibitem [{\citenamefont {{Batygin}}\ and\ \citenamefont
  {{Brown}}(2016)}]{2016AJ....151...22B}%
  \BibitemOpen
  \bibfield  {author} {\bibinfo {author} {\bibfnamefont {K.}~\bibnamefont
  {{Batygin}}}\ and\ \bibinfo {author} {\bibfnamefont {M.~E.}\ \bibnamefont
  {{Brown}}},\ }\href {\doibase 10.3847/0004-6256/151/2/22} {\bibfield
  {journal} {\bibinfo  {journal} {\aj}\ }\textbf {\bibinfo {volume} {151}},\
  \bibinfo {eid} {22} (\bibinfo {year} {2016})},\ \Eprint
  {http://arxiv.org/abs/1601.05438} {arXiv:1601.05438 [astro-ph.EP]}
  \BibitemShut {NoStop}%
\bibitem [{\citenamefont {Scholtz}\ and\ \citenamefont
  {Unwin}(2019)}]{Scholtz:2019csj}%
  \BibitemOpen
  \bibfield  {author} {\bibinfo {author} {\bibfnamefont {J.}~\bibnamefont
  {Scholtz}}\ and\ \bibinfo {author} {\bibfnamefont {J.}~\bibnamefont
  {Unwin}},\ }\href@noop {} {\  (\bibinfo {year} {2019})},\ \Eprint
  {http://arxiv.org/abs/1909.11090} {arXiv:1909.11090 [hep-ph]} \BibitemShut
  {NoStop}%
\bibitem [{\citenamefont {Kolb}\ and\ \citenamefont
  {Tkachev}(1993)}]{Kolb:1993zz}%
  \BibitemOpen
  \bibfield  {author} {\bibinfo {author} {\bibfnamefont {E.~W.}\ \bibnamefont
  {Kolb}}\ and\ \bibinfo {author} {\bibfnamefont {I.~I.}\ \bibnamefont
  {Tkachev}},\ }\href {\doibase 10.1103/PhysRevLett.71.3051} {\bibfield
  {journal} {\bibinfo  {journal} {Phys. Rev. Lett.}\ }\textbf {\bibinfo
  {volume} {71}},\ \bibinfo {pages} {3051} (\bibinfo {year} {1993})},\ \Eprint
  {http://arxiv.org/abs/hep-ph/9303313} {arXiv:hep-ph/9303313} \BibitemShut
  {NoStop}%
\bibitem [{\citenamefont {Levkov}\ \emph {et~al.}(2018)\citenamefont {Levkov},
  \citenamefont {Panin},\ and\ \citenamefont {Tkachev}}]{Levkov:2018kau}%
  \BibitemOpen
  \bibfield  {author} {\bibinfo {author} {\bibfnamefont {D.}~\bibnamefont
  {Levkov}}, \bibinfo {author} {\bibfnamefont {A.}~\bibnamefont {Panin}}, \
  and\ \bibinfo {author} {\bibfnamefont {I.}~\bibnamefont {Tkachev}},\ }\href
  {\doibase 10.1103/PhysRevLett.121.151301} {\bibfield  {journal} {\bibinfo
  {journal} {Phys. Rev. Lett.}\ }\textbf {\bibinfo {volume} {121}},\ \bibinfo
  {pages} {151301} (\bibinfo {year} {2018})},\ \Eprint
  {http://arxiv.org/abs/1804.05857} {arXiv:1804.05857 [astro-ph.CO]}
  \BibitemShut {NoStop}%
\bibitem [{\citenamefont {Eggemeier}\ and\ \citenamefont
  {Niemeyer}(2019)}]{Eggemeier:2019jsu}%
  \BibitemOpen
  \bibfield  {author} {\bibinfo {author} {\bibfnamefont {B.}~\bibnamefont
  {Eggemeier}}\ and\ \bibinfo {author} {\bibfnamefont {J.~C.}\ \bibnamefont
  {Niemeyer}},\ }\href {\doibase 10.1103/PhysRevD.100.063528} {\bibfield
  {journal} {\bibinfo  {journal} {Phys. Rev. D}\ }\textbf {\bibinfo {volume}
  {100}},\ \bibinfo {pages} {063528} (\bibinfo {year} {2019})},\ \Eprint
  {http://arxiv.org/abs/1906.01348} {arXiv:1906.01348 [astro-ph.CO]}
  \BibitemShut {NoStop}%
\bibitem [{\citenamefont {Levkov}\ \emph {et~al.}(2020)\citenamefont {Levkov},
  \citenamefont {Panin},\ and\ \citenamefont {Tkachev}}]{Levkov:2020txo}%
  \BibitemOpen
  \bibfield  {author} {\bibinfo {author} {\bibfnamefont {D.}~\bibnamefont
  {Levkov}}, \bibinfo {author} {\bibfnamefont {A.}~\bibnamefont {Panin}}, \
  and\ \bibinfo {author} {\bibfnamefont {I.}~\bibnamefont {Tkachev}},\
  }\href@noop {} {\  (\bibinfo {year} {2020})},\ \Eprint
  {http://arxiv.org/abs/2004.05179} {arXiv:2004.05179 [astro-ph.CO]}
  \BibitemShut {NoStop}%
\end{thebibliography}%

\end{document}